\documentclass[aps,twocolumn,amsmath,amssymb]{revtex4}
\usepackage{graphicx}% Include figure files
\usepackage{dcolumn}% Align table columns on decimal point
\usepackage{bm}% bold math
\usepackage{amssymb}
\usepackage{amsmath}
\begin{document}

%\preprint{APS}

\title{STRUCTURED THALAMOCORTICAL CONNECTIVITY
REVEALED BY RANDOM WALKS ON COMPLEX NETWORKS}

\author{Luciano da Fontoura Costa} 
\email{luciano@if.sc.usp.br}
 \affiliation{Instituto de F\'{\i}sica de S\~{a}o Carlos, Universidade
 de S\~{a}o Paulo, Av. Trabalhador S\~{a}o Carlense 400, Caixa Postal
 369, CEP 13560-970, S\~{a}o Carlos, S\~ao Paulo, Brazil}
\author{Olaf Sporns}\email{osporns@indiana.edu}
  \affiliation{Department of Psychological and Brain Sciences, Indiana 
  University, Bloomington, IN 47405, USA}

\date{5th Feb 2006}

\begin{abstract}
The segregated regions of the mammalian cerebral cortex and thalamus
form an extensive and complex network, whose structure and function
are still only incompletely understood. The present article describes
an application of the concepts of complex networks and random walks
that allows the identification of non-random, highly structured
features of thalamocortical connections, and their potential effects
on dynamic interactions between cortical areas in the cat brain.
Utilizing large-scale anatomical data sets of this thalamocortical
system, we investigate uniform random walks in such a network by
considering the steady state eigenvector of the respective stochastic
matrix.  It is shown that thalamocortical connections are organized in
such a way as to guarantee strong correlation between the outdegree
and occupancy rate (a stochastic measure potentially related to
activation) of each cortical area.  Possible organizational principles
underlying this effect are identified and discussed.

\end{abstract}


\maketitle

The relationship between the topology of structural connection
patterns and cortical dynamics currently represents a significant
challenge to brain theory.  Numerous neuroanatomical studies have
revealed that the pathways of the mammalian thalamocortical system
exhibit specific patterns ranging in scale from interconnections
linking whole brain regions to intra-areal patterns of connections
between cell populations or individual cortical neurons.
Comprehensive descriptions of large-scale anatomical patterns of
cortical connectivity have been collated for several mammalian species
(e.g. \cite{Felleman_Essen:1991,Scannell_etal:1999}).  Quantitative
analysis has revealed that these patterns are neither completely
regular nor completely random \cite{Sporns_etal:2004}, but exhibit
specific attributes, such as short path lengths combined with high
clustering \cite{Hilgetag_etal:2000, Sporns_etal:2000}, short wiring
and distinctive hierarchical features \cite{Costa_Sporns:2006}.

The current article proposes a novel macroscopic approach to studying
neuronal dynamics in cortical maps based on concepts from random walks
(e.g.~\cite{Spitzer:2001,Avraham:2000}) and complex networks
(e.g.~\cite{Newman:2003,Albert_Barab:2002,Boccaletti_etal:2006}).
Specifically, we examine the interactions between cortex and thalamus
in the mammalian central nervous system.  The cerebral cortex consists
of a network of interconnected functionally specialized regions, each
of which maintains reciprocal connections with a specific set of
thalamic nuclei.  These nuclei relay cortical inputs and outputs, and
are thought to play a role in coordinating cortical interactions. The
thalamocortical architecture is represented in terms of a complex
network, obtained by assigning a node to each cortical region and
thalamic nucleus (nodes) and distributing anatomical connections as
links (or edges) between such nodes.  The flow of activation of each
node is then modeled by an agent engaged in a random walk on the edges
of the network.  The movement of the agent expresses the exchange or
flow of information between adjacent nodes, analogous to a diffusive
process.  After a long number of time steps $T$, the occupancy rate,
i.e. the number of times each node was visited divided by $T$,
captures the rate of participation of each node in network-wide
information flow, a dynamic measure analogous to a cortical activation
level. In order to relate this dynamical measure to the underlying
network topology, we examine the outdegree of each network node.

The anatomical connection matrix of cat cortex and thalamus was
obtained from the study by Scannell et al.~\cite{Scannell_etal:1999}
and contained 53 cortical regions and 42 thalamic regions (Fig. 1a).
Most corticocortical connections and all thalamocortical connections
are reciprocal (i.e. the matrix is nearly symmetric), and no
connections exist between any of the thalamic nodes.  Following the
analysis in~\cite{Hilgetag_etal:2000} the 53 cortical areas can be
divided into four functionally distinct clusters: visual (17, 18, 19,
PLLS, PMLS, ALLS, AMLS, VLS, DLS, 21a, 21b, 20a, 20b, 7, AES, PS),
auditory (AI, AII, AAF, P, VPc, EPp, Tem), somatomotor (31, 3b, 1, 2,
SII, SIV, 4g, 4, 6l, 6m, 5Am, 5Al, 5Bm, 5Bl, SSAi, SSAo), and
frontolimbic (PFCMil, PFCMd, PFCL, Ia, Ig, CGa, CGp, RS, 35, 36, pSb,
Sb, Enr, Hipp).  Visual and auditory clusters are combined into a
`posterior' cluster, while somatomotor and frontolimbic clusters are
combined into an `anterior' cluster.  The adjacency matrix is
henceforth represented as $K$, with the presence of a directed
connection from region $j$ to $i$ indicated as $K(i,j)=1$.

Because of the relatively small number of involved regions and the
fact that any node can be reached from any other node in the network
(i.e. it forms a single connected component), it is possible to
characterize the steady state node occupation effectively in terms of
the dominant eigenvector of the respective stochastic matrix (the
random walk can be understood as a random driven Markov chain)
\cite{Bremaud:2001}.  In order to do so, a stochastic version $S$ of
the adjacency matrix $K$ is obtained by making

\begin{eqnarray}
  O(j) = \sum_{i=1}^{N} K(i,j)  \\
  S(i,j) = K(i,j)/O(j) 
\end{eqnarray}

where $O(j)$ is the \emph{outdegree} of node $j$, i.e. the number of
edges emanating from that node. Note that $S$ is a stochastic matrix
because $S(i,j) \ge 0$ for any $i$ and $j$ and $\sum_{i=1}^{N}S(i,j)=1$
for any $j$.  At steady state, the occupation rate of the random walk
movements can be obtained from the eigenvector equation $S \vec{v} =
\vec{v}$, where $\vec{v}$ is the eigenvector of the stochastic matrix 
$S$ associated to the unit eigenvalue~\cite{Bremaud:2001}.  The
eigenvector $\vec{v}$ is normalized so that $\sum_{i=1}^{N} v_i = 1$.
The average occupancy rate for each cortical region $i$ is therefore
given by the eigenvector element $\vec{v}(i)$.

Figure 1b shows the outdegree ($O$) versus the occupancy rate ($R$)
for the cortical regions obtained by the eigenvalue analysis
considering \emph{all} thalamocortical interconnections.  Posterior
and anterior regions are identified by `$\times$' and `$\ast$'
respectively.  This network exhibits a pronounced positive correlation
(Pearson correlation r = 0.84) between the topological and dynamical
measures.  We examined the hypothesis that this strong correlation is
due to the specific organizational pattern of thalamocortical
connections.  First, we repeat the analysis considering \emph{only}
the cortico-cortical connections.  Now, a markedly weaker correlation
($r = 0.54$) was observed between the topological (i.e. outdegree) and
dynamical (i.e. occupancy rate) measures, which are illustrated in
Figure 1c.  When replacing the thalamocortical connections by the same
number of randomly distributed edges (maintaining reciprocity), we
again obtained a weaker correlation ($r = 0.65 \pm 0.044$ considering
20 random realizations) between outdegree and occupancy ratio.
Randomization of the corticocortical connections while keeping the
original thalamocortical edges results in a decrease, though slightly
less marked, of the correlation between outdegree and occupancy rates
($r = 0.67 \pm 0.041$ for 20 random realizations).  Similar figures
were obtained regarding the correlation between indegree and
activation rate.  Such results seem to indicate that the specific
pattern of thalamocortical connections promotes a strong correlation
between topological and dynamical features of cortical interactions.

In order to try to infer which structural features of the
thalamocortical interconnections are principally responsible for
generating the correlation between topology and dynamics, each
thalamic node was isolated together with the respective in and
outbound connections -- defining a \emph{V-structure}, as allowed by
the complete lack of interconnectivity between these nodes. The
shortest paths between each of the cortical regions at the destination
of the outbound edges of thalamic node $i$ and all the cortical
regions at the origin of the inbound edges are computed. The
respective average (Av) and standard deviation values (St) are
calculated and shown in Figure 1d in decreasing order.  The plot shows
that the additional edges (and therefore outdegree) imposed by the
thalamic V-structures tend to implement cycles of just a few edges
among the group of cortical regions to which it is attached. In other
words, if the shortest path between two cortical areas attached to a
V-structure involves $k$ edges, the connection to the thalamic
structure will define a cycle with $k+2$ edges.  This effect
contributes to a directly proportional increase of the occupancy rate,
being potentially related to the observed correlation between the
considered topological and dynamical features.  Other elements of the
thalamocortical structure also contributing to the strong correlation
include the marked similarity between the in and outdegree at each
node as well as the fact that the whole network defines a connected
component.

In conclusion, we have shown in this article that a novel combination
of concepts from complex networks and random walks/Markov chains can
reveal important structural properties underlying thalamocortical
architecture in the cat brain.  Our analysis demonstrates the presence
of a strong correlation between network topology (in and outdegree)
and dynamic features of cortical activation (occupancy rate) as
modeled by random walks.  The basic feature of the thalamic
connections that is responsible for establishing such a correlation
has also been investigated in terms of shortest path analysis between
cortical nodes participating in each of the thalamic V-structures.  A
possible critical feature of such circuits was identified as the
reliatively short paths between subgroups of cortical regions.  In a
na\"{\i}ve analogy, it is as if the thalamic V-structures provided a
feedback mirror to groups of closely connected cortical regions.  As
for the possible implications of the correlation between outdegree and
occupancy rate, it has the two following interesting effects: (1) hubs
of connectivity are likely to be hubs of neuronal information flow and
activity; and (2) in the case of scale free topology (degrees), the
dynamic participation or cortical activation at each node will also
become scale free.

The reported findings have potentially important implications for the
understanding of cortical functional architecture and pave the way to
a number of subsequent studies, including the comparison of cortical
connectivity of other species and the use of other kinds of random
walks and dynamics.  At the same time, several other problems in
complex networks can be addressed in terms of the correlation
measurement proposed in this article.

\begin{acknowledgments}

Luciano da F. Costa thanks CNPq (308231/03-1) for sponsorship.  Olaf
Sporns was supported by the James S. McDonnell Foundation.

\end{acknowledgments}

%\pagebreak

\begin{figure*}
\begin{center}
\hspace{0.4cm}
\includegraphics[scale=0.23]{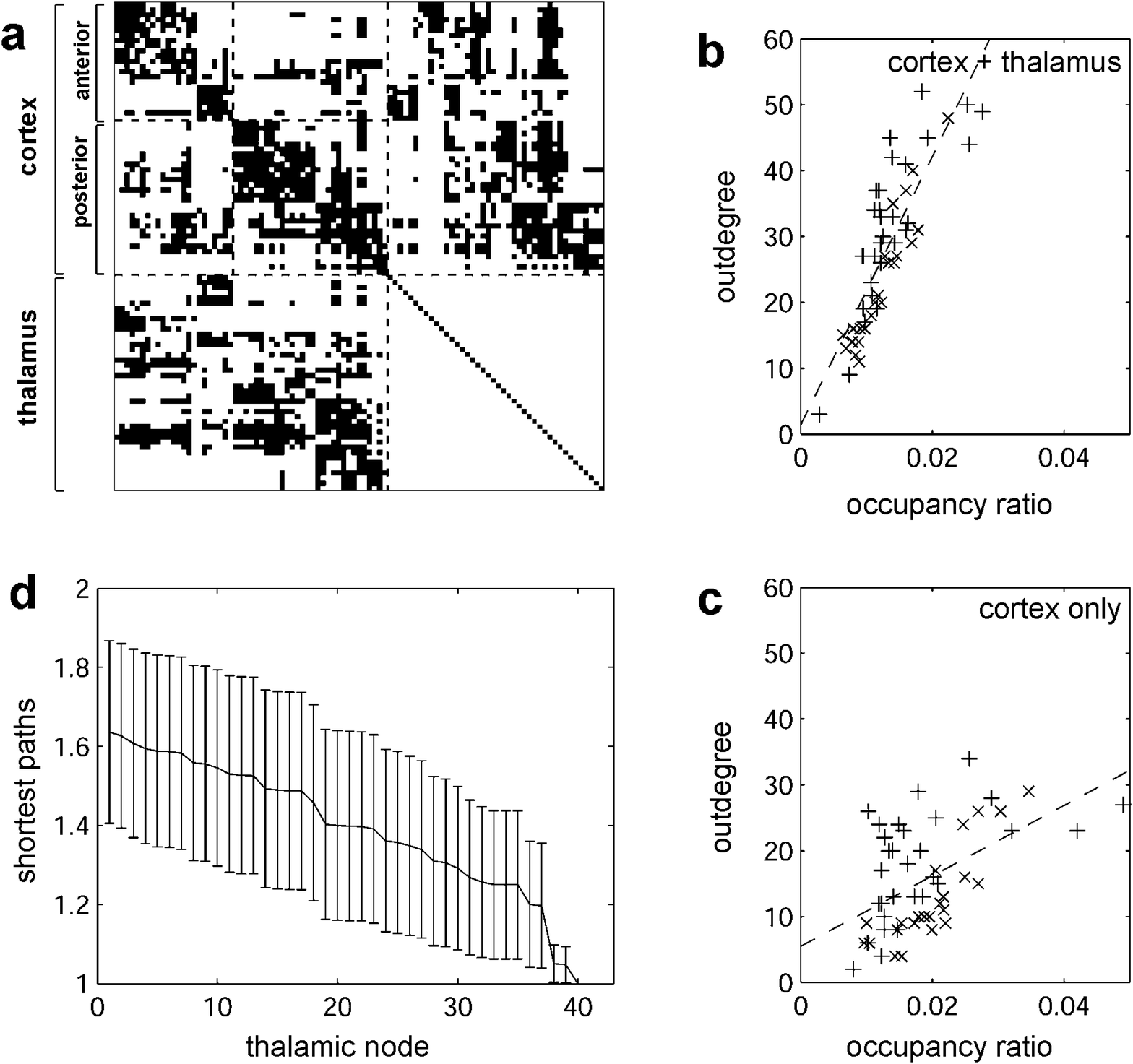}
\end{center}
\caption{\label{scatter} Figure 1.}
\end{figure*}

%\pagebreak 

\vspace{4cm}

List of Captions[o3]:

Adjacency matrix of the directed complex network defined by the cat
thalamocortical connections (a).  The scatterplot of outdegree ($O$)
and occupancy rate ($R$) considering all thalamocortical connections
(b) and only cortical connections (c).  The average $\pm$ standard
deviation of the shortest paths between the cortical nodes connecting
to thalamic nodes, considering only cortical interconnections (d).

%\pagebreak

\bibliography{apl_thal}% Produces the bibliography via BibTeX.

\end{document}